\documentclass{ptephy_v1}

\usepackage{ulem,bm}

\begin{document}

\title{Single-particle decomposition of nuclear surface diffuseness}

\author{Wataru Horiuchi}
\affil[1]{Department of Physics,
  Hokkaido University, Sapporo 060-0810, Japan
  \email{whoriuchi@nucl.sci.hokudai.ac.jp}}

\begin{abstract}
  Nuclear surface diffuseness reflects spectroscopic
  information near the Fermi level.
  I propose a way to decompose the surface diffuseness into
  single-particle (s.p.) contributions in a quantitative way.
  Systematic behavior of the surface diffuseness
  of neutron-rich even-even O, Ca, Ni, Sn, and Pb isotopes
  is analyzed with a phenomenological mean-field approach.
  The role of the s.p. wave functions near the Fermi level is explored:
  The nodeless s.p. orbits form a sharp nuclear surface, while
  the nodal s.p. orbits contribute to diffusing
  the nuclear surface.
\end{abstract}

\subjectindex{xxxx, xxx}

\maketitle
\section{Introduction}

Studying the structure of short-lived neutron-rich unstable nuclei
is one of the most important subjects in nuclear physics.
Various exotic phenomena such as halos~\cite{Tanihata85, Ozawa01, Tanihata13, Bagchi20},
neutron skin~\cite{Suzuki95}, and deformation~\cite{Takechi12, Takechi14},
have been found through systematic studies of nuclear radii.
Since the nuclear radius is sensitive to the nuclear density
profile around the nuclear surface,
these drastic nuclear structure changes near the Fermi level
can be deduced from measurements
of the interaction or total reaction cross section.
For a deeper understanding of the structure of unstable nuclei
predicted by theoretical models,
one needs to know more details on the density profiles
near the nuclear surface, which may include
more information on spectroscopic properties
other than the nuclear radius.

To know the whole density distribution,
electron scattering has been used to extract the charge distribution
of stable nuclei~\cite{deVries87}.
Combining known charge or proton density distributions,
the neutron density distribution has been deduced
for some stable nuclei by measuring proton elastic scattering
cross sections up to large scattering
angles~\cite{Terashima08, Zenihiro10, Sakaguchi17}.
The electron scattering~\cite{scrit2017}
and proton-elastic scattering~\cite{Matsuda13,Chebotaryov18}
measurements have been extended for unstable nuclei.
However, it is still hard to determine whole density distributions
as it requires cross section measurement up to large scattering angles.
Even though if all the nuclear density distributions cannot be determined,
the nuclear surface ``diffuseness'',
which quantifies the nuclear surface thickness,
is a promising measure of the surface density profile and can be deduced
accurately using nucleus-proton scattering as demonstrated
in Ref.~\cite{Hatakeyama18}.
To get the surface diffuseness, one only needs the cross sections
at forward angles up to the first peak position,
which is advantageous to study unstable nuclei
by the inverse kinematics.

Understanding how the surface diffuseness forms
is of particular importance as the next order information of the nuclear radius.
Evaluating the surface diffuseness as well as the neutron skin thickness
also impacts on constraining the equation of state of nuclear matter
from finite nuclear systems~\cite{Reinhard10,Chen10,Iida04,Horiuchi17}.
Thus far, various phenomena have been observed
which may be closely related to the surface diffuseness:
Kink structures of total reaction or interaction cross section
in neutron-rich Ne~\cite{Takechi12} and Mg~\cite{Takechi14} isotopes
due to nuclear deformation~\cite{Minomo11,Minomo12,Sumi12,Horiuchi12,Watanabe14,Horiuchi15},
and core swelling phenomena in spherical Ca isotopes~\cite{Tanaka20,Horiuchi20}.
The surface diffuseness changes at the major or subshell
were pointed out in Ref.~\cite{Hatakeyama18}.
A systematic and extensive theoretical study
on the surface diffuseness was given in Ref.~\cite{Scamps13}.
Bubble nuclei, which have depleting central density
due to the lack of the occupation of the $s$-wave orbit,
can be identified by measuring the surface diffuseness~\cite{Choudhary20}.
The surface diffuseness is drastically
enhanced when the $pf$ mixed orbits are occupied in
the island of inversion in the neutron-rich Ne and Mg
isotopes~\cite{Choudhary21}.

The density profile near the nuclear surface
could mainly be formed by the single-particle (s.p.)
wave functions near the Fermi level.
The purpose of this paper is to clarify the role of
the s.p. wave functions in forming the nuclear surface.
I propose a practical and convenient way
to decompose the surface diffuseness into
each s.p. contribution. As a first step, for simplicity,
I only consider spherical configurations generated from
a phenomenological mean-field potential.

The paper is organized as follows.
Section~\ref{models.sec} describes theoretical models
used in this paper.
Section~\ref{formulation.sec} define the nuclear surface diffuseness
based on a familiar two-parameter Fermi (2pF) density distribution.
I derive a relationship between these diffuseness and radius parameters
by relating the 2pF density with one-body density
generated from a mean-field model, allowing one to
decompose the total surface diffuseness into s.p. contributions.
Section~\ref{WS.sec} describes a phenomenological mean-field model
employed in this paper. The validity of this approach is confirmed
by a comparison of available experimental data related to nuclear size, i.e.,
the interaction cross section and charge radius.
Section~\ref{diff.sec} shows the total surface diffuseness obtained
from the present approach and overviews its general behavior.
Section~\ref{results.sec} is devoted to
a detailed analysis and discusses the role of s.p.
orbits in the total surface diffuseness for each isotope.
In Secs.~\ref{O.sec}--\ref{Pb.sec}, results
for O, Ca and Ni, Sn, and Pb isotopes are respectively discussed.
Section~\ref{global.sec} explores a global feature of
the s.p. wave functions by taking examples of the neutron dripline nuclei.
The role of the s.p. wave functions near the Fermi level is clarified
through an analysis of the radii of the s.p. orbits.
Finally, conclusion is given in Sec.~\ref{conclusion.sec}

\section{Theoretical models}
\label{models.sec}

\subsection{Single-particle decomposition of nuclear surface diffuseness}
\label{formulation.sec}

\begin{figure}[htb]
  \centering
\includegraphics[width=12cm]{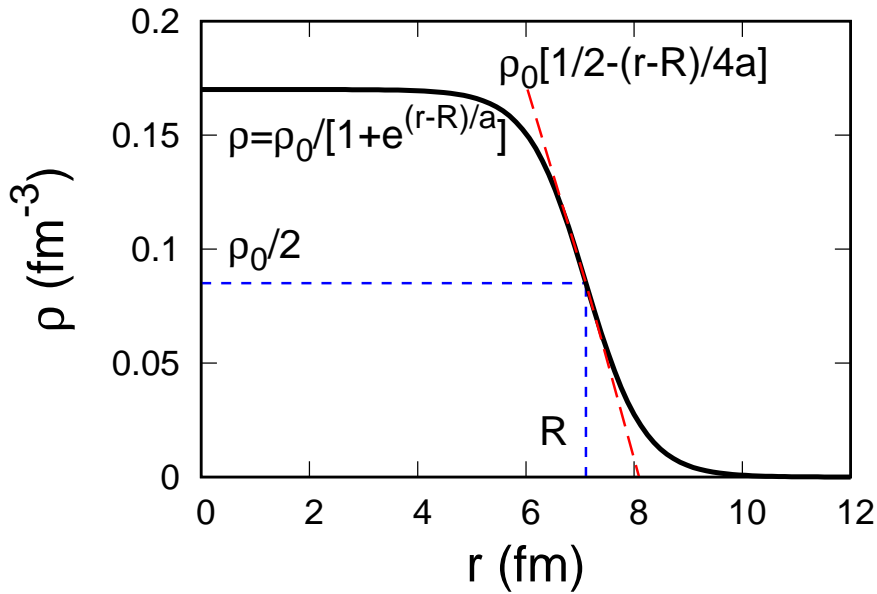}      
\caption{Geometry of a 2pF density distribution
  for $^{208}$Pb with $\rho_0=0.17$ fm$^{-3}$,
  $R=1.2\times 208^{1/3}$ fm, and $a=0.54$ fm~\cite{BM}.
  The dashed line indicates the slope of the 2pF distribution
  at the half density ($\rho_0/2$) radius $R$ which are indicated
  by the dotted lines.
}
\label{2pF.fig}
\end{figure}

To define the nuclear surface diffuseness or thickness parameter,
I employ a two-parameter Fermi (2pF) function
as an anzatz of the density distribution of a nucleus
\begin{align}
  \rho_{\rm 2pF}(r)=\frac{\rho_0}{1+\exp\left[(r-R)/a \right]}.
  \label{2pF.eq}
\end{align}
For given radius parameter $R$
and diffuseness parameter $a$,
$\rho_0$ can be determined uniquely
by the normalization condition $4\pi \int_0^\infty dr\, r^2\rho_{\rm 2pF}(r)=A$
with $A$ being the mass number of a nucleus.
Expanding the above distribution at $R$
and taking the first-order term, I get
\begin{align}
  \rho_{\rm 2pF}(R)+(r-R)\left.\frac{d\rho_{\rm 2pF}}{dr}\right|_{r=R}=
  \rho_0\left(\frac{1}{2}-\frac{r-R}{4a}\right).
\label{2pFderi.eq}
\end{align}
Figure~\ref{2pF.fig} illustrates geometry of a typical 2pF distribution
for $^{208}$Pb with a standard parameter $\rho_0=0.17$ fm$^{-3}$,
$R=1.2\times 208^{1/3}$ fm, and $a=0.54$ fm~\cite{BM}.
This clearly shows that the diffuseness parameter $a$ corresponds
to the slope of the 2pF distribution at the radius parameter $R$,
where the 2pF distribution becomes approximately half of the
central density $\rho_{\rm 2pF}(0)\approx \rho_0$.
This 2pF-type density 
  well approximates typical density distributions~\cite{Hatakeyama18}.
  Because of its simplicity this first-order approximation,
which corresponds to the trapezoidal distribution with the surface
thickness $4a$, was applied to correct
the sharp nuclear surface of the black-sphere model
for high-energy proton-nucleus scattering~\cite{Kohama16}.
I remark that the 2pF distribution may not describe a weakly bound neutron tail.
The limitation of this sort of density distribution, especially for
halo nuclei, was discussed in Ref.~\cite{Mizutori00}.

Decomposition of the nuclear density distribution
can be made by assuming that the total density is composed of the sum of
single-particle (s.p.) density distributions  $\bar{\rho}_i$ as
  \begin{align}
    \rho(r)=\sum_{i=1}^A\bar{\rho}_i(r).
\label{spdens.eq}
  \end{align}
Given the relationships of Eqs.~(\ref{2pF.eq}) and (\ref{2pFderi.eq}),
it is straightforward to decompose the surface diffuseness
parameter into each s.p. contribution.
Differentiating this density of Eq.~(\ref{spdens.eq}) at $r=R$,
I get a relation by assuming $\rho(r)=\rho_{\rm 2pF}(r)$:
  \begin{align}
    \left.\frac{d\rho}{dr}\right|_{r=R}=\sum_{i=1}^AD_i=-\frac{\rho_0}{4a}
  \end{align}
  with
  \begin{align}
    D_i=\left.\frac{d\bar{\rho}_i}{dr}\right|_{r=R}
\label{di.eq}    
  \end{align}
  Finally, the surface diffuseness is expressed by
  \begin{align}
    a=-\frac{\rho_0}{4}\left(\sum_{i=1}^A D_i\right)^{-1}.
\label{directa.eq}
  \end{align}
  The diffuseness parameter $a$ is inversely proportional
  to the sum of $D_i$.
   To evaluate $D_i$,  the $R$ value should be defined
    for general density distributions. As shown later, the $a$ value
    can reasonably be obtained with a proper choice of $R$.
  I will investigate these $D_i$ values for each s.p. density distribution
  and quantify the s.p. contribution to the total surface diffuseness $a$.
  
\subsection{Density distributions of O, Ca, Ni, Sn, and Pb isotopes}
\label{WS.sec}

In this paper, I generate the s.p. density distributions $\bar{\rho}_i$
from a phenomenological mean-field model~\cite{BM} for the sake of simplicity.
The mean-field potential is parameterized as
\begin{align}
  V(r)=V_0f(r)+V_1(\bm{l}\cdot\bm{s})r_0^2\frac{1}{r}\frac{df}{dr}
  +V_C(r)\frac{1}{2}(1-\tau_3)
\end{align}
with the Woods-Saxon form factor,
$f(r)=\left\{1+\exp\left[(r-R_{\rm WS})/a_{\rm WS}\right]\right\}^{-1}$
and
\begin{align}
  V_0=-51+33\frac{N-Z}{A}\tau_{3}, \quad V_1=22- 14\frac{N-Z}{A}\tau_{3}
\end{align}
in unit of MeV
with $N$ and $Z$ being the neutron and proton numbers, respectively,
and $\tau_3=\pm 1$ corresponding to $+$ ($-$) for neutron (proton).
The radius and diffuseness parameters of the potential are given respectively by
$R_{\rm WS}=1.27A^{1/3}$ fm and $a_{\rm WS}=0.67$ fm.
The Coulomb potential term $V_C$ is taken
as a uniform charge distribution with radius $R_{\rm WS}$.
I calculate all s.p. bound states and obtain their radial s.p. wave functions
$\psi_{nlj}(r)$ with $n$, $l$, $j$ being radial, orbital, and angular momentum
quantum numbers, respectively.
By averaging over magnetic quantum numbers
and integrating over the spin coordinate,
the density distributions are expressed by a sum of
squared radial s.p. wave functions,
in which the outermost s.p. level is averaged and filled equally as
\begin{align}
  \rho(r)=\sum_{n,l,j\in {\rm occ.}}(2j+1)|\psi_{nlj}(r)|^2+N_{v}|\psi_{n_vl_vj_v}(r)|^2
\end{align}
with $4\pi \int_0^\infty dr\, r^2 |\psi_{nlj}(r)|^2=1$
and $4\pi \int_0^\infty dr\, r^2 \rho(r)=A$,
where $N_v$ is the number of the outermost nucleon
with the quantum numbers $n_v, l_v$, and $j_v$.
Note that $\bar{\rho}_i$ in Eq.~(\ref{spdens.eq})
is nothing but $|\psi_{nlj}|^2$.
The rms radius of each s.p. orbit can be evaluated by
\begin{align}
  r_{nlj}=\sqrt{4\pi\int_0^\infty dr\,r^4|\psi_{nlj}(r)|^2}.
\end{align}
and the rms matter radius is given
by
\begin{align}
  R_m=\sqrt{\frac{4\pi}{A}\int_0^\infty dr\,r^4 \rho(r)}.
\label{radm.eq}
\end{align}

I calculate such density distributions
for proton magic nuclei for O, Ca, Ni, Sn, and Pb isotopes.
Since the parameter set of Ref.~\cite{BM} was determined
to reproduce the properties of the medium- to heavy-mass nuclei,
this parameter set may not be appropriate to
describe light nuclei such O isotopes~\cite{Horiuchi07, Ibrahim09}.
In fact, the neutron dripline of O isotopes
is predicted at $^{28}$O,
which contradicts the measurements~\cite{Langevin85,Guillemaud90,Tarasov97}.
Therefore, I modify the potential parameter for O isotopes
as $R_{\rm WS}=1.20A^{1/3}$ fm and $a_{\rm WS}=0.60$ fm to reproduce the neutron
dripline $^{24}$O. For the other isotopes, I use the original parameter sets
as they are.
Finally, I find even-even isotopes,
$^{14-24}$O, $^{34-60}$Ca, $^{50-86}$Ni, $^{96-164}$Sn, and $^{178-266}$Pb.
I remark that some exotic nuclear structure was predicted in the neutron-rich
Ca isotopes for $N>40$ that must be reflected
in the surface diffuseness~\cite{Meng02,Hagen12,Hagen13,Hove18};
however, this parameter set does not produce
a bound $0g_{9/2}$ orbit for $N>40$.
A detailed study along the isotopic chain beyond $N=40$
using a more realistic structure model
is quite interesting and will be reported elsewhere.

To evaluate the validity of calculated density distributions,
the nuclear radii are useful observables as they properly reflect
the density profile near the nuclear surface.
First, I calculate the total reaction cross section at high-incident energy,
which directly reflects the matter density profile.
The nucleon-target profile functions~\cite{NTG}
in the Glauber model~\cite{Glauber} is employed,
allowing us to get reliable total reaction cross sections
of high-energy nucleus-nucleus collisions. 
Inputs of the model are the density distributions
of the projectile and target nuclei, and the so-called profile function,
which describes nucleon-nucleon scattering.
The parameter of the profile function is taken from Ref.~\cite{Ibrahim08}.
This model nicely describes the total reaction cross sections
at high incident energies. See
Refs.~\cite{Horiuchi06,Horiuchi10, Horiuchi12, Horiuchi15, Horiuchi16, Horiuchi17b,Nagahisa18}
for various applications.

\begin{figure}[htb]
  \centering
  \includegraphics[width=0.7\linewidth]{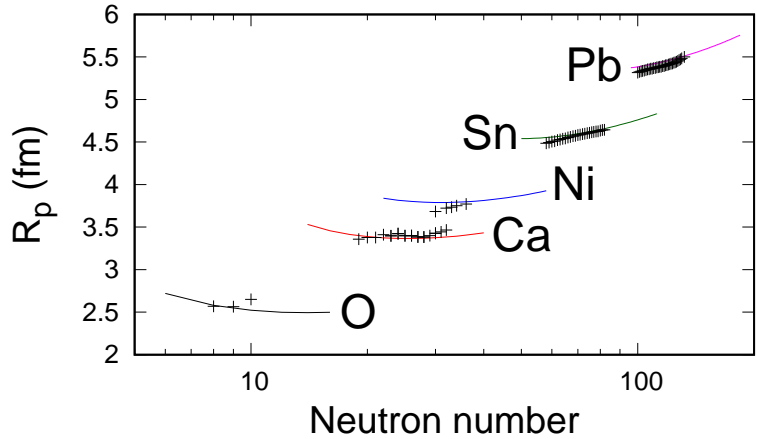}      
\caption{Rms point-proton radii of even-even O, Ca, Ni, Sn, and Pb isotopes
  as a function of neutron number.
  The experimental point-proton radii,
  which are indicated by plus symbols,
  are extracted from the experimental charge radii~\cite{Angeli13,Ruiz16}.}
\label{radii_p.fig}
\end{figure}

Table~\ref{rcs.tab} compares the calculated total reaction
and experimental interaction cross-section data on
a carbon target at high incident energies, 1000 MeV/nucleon for O isotopes
and 300 MeV/nucleon for Ca isotopes.
Good agreement for all listed O and Ca isotopes
including the recent cross-section data
for neutron-rich Ca isotopes~\cite{Tanaka20} is attained.
The calculated root-mean-square (rms) matter radii are also consistent with
these obtained in Ref.~\cite{Ozawa01b,Chulkov96,Tanaka20}.
In addition to the cross-section data,
I also compare in Fig.~\ref{radii_p.fig}
the rms point-proton radii of O, Ca,
Ni, Sn, and Pb isotopes
evaluated by $R_p=\sqrt{\frac{4\pi}{Z}\int_0^\infty dr\, r^4 \rho_p(r)}$,
where $\rho_p(r)$ is the proton density.
The ``experimental'' point-proton radii
are extracted from experimental charge radii~\cite{Angeli13,Ruiz16}.
Overall agreement with the theory and experimental data is obtained,
and thus the density distributions employed in this paper
reasonably describe the density profile near the nuclear surface.

\begin{table}[ht]
\begin{center}
  \caption{Calculated rms matter radii and
    comparison of calculated total reaction and experimental
    interaction cross sections of O and Ca isotopes on a carbon target.
  The cross sections are computed at 1000 and 300 MeV/nucleon
  for the O and Ca isotopes, respectively.
  Experimental data is taken from Refs.~\cite{Ozawa01b,Chulkov96,Tanaka20}.}
\begin{tabular}{cccc}
\hline\hline
    &$R_m$ (fm)&$\sigma_R$ (mb)& $\sigma_I$(Expt.) (mb)\\
\hline
$^{16}$O &2.55&980 & 982$\pm  6$\\
$^{18}$O &2.69&1060&1032$\pm 26$\\
$^{20}$O &2.83&1140&1078$\pm 10$\\
$^{22}$O &2.95&1210&1172$\pm 22$\\
$^{24}$O &3.22&1370&1318$\pm 52$\\
\hline
$^{42}$Ca&3.40&1430&1463$\pm 13 \pm 6$\\
$^{44}$Ca&3.46&1460&1503$\pm 12 \pm 6$\\
$^{46}$Ca&3.51&1500&1505$\pm  9 \pm 6$\\
$^{48}$Ca&3.57&1540&1498$\pm 17 \pm 6$\\
$^{50}$Ca&3.65&1600&1615$\pm 42 \pm 7$\\
\hline\hline 
\end{tabular}
\end{center}
\label{rcs.tab}
\end{table}

\subsection{Evaluation of surface diffuseness from single-particle densities}
\label{diff.sec}

\begin{figure}[ht]
\begin{center}
\includegraphics[width=\linewidth]{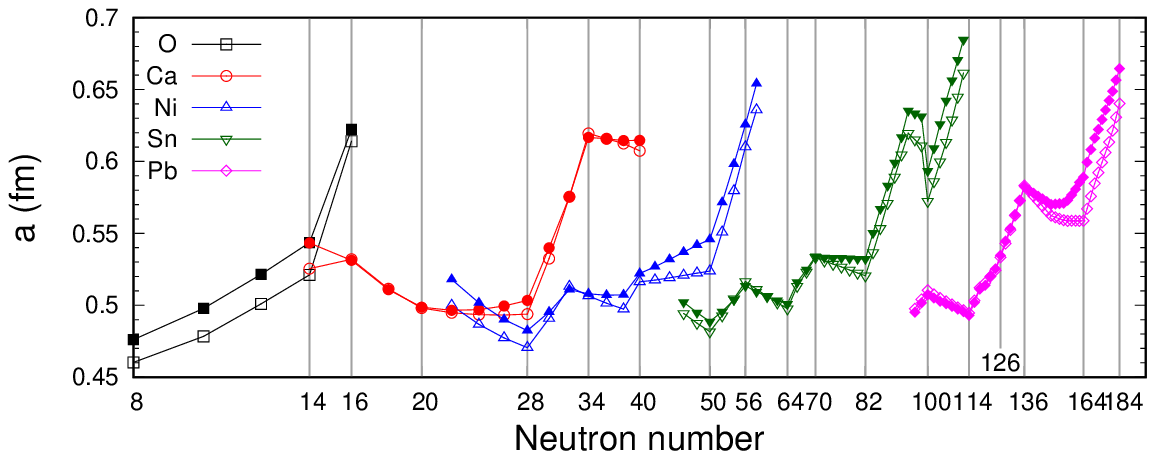}      
\caption{Diffuseness parameters 
     of even-even O, Ca, Ni, Sn, and Pb isotopes
     evaluated from Eq.~(\ref{directa.eq}) as a function of neutron number.
     Closed symbols denote those obtained with
     the minimization of Eq.~(\ref{mindens.eq}).
    Thin lines connecting symbols are a guide to the eye.
    Note that the neutron number is given in logarithmic scale and
    some magic and semi-magic numbers are indicated by
    thin vertical lines.}
 \label{diff.fig}
\end{center}
\end{figure}

I have confirmed that the calculated density distributions
reasonably reproduce the
existing experimental data related to the nuclear size.
In this subsection, I extract the diffuseness parameter $a$
from those density distributions by using
the relation of Eq. (\ref{directa.eq})
with the reference radius parameter $R$.
As was demonstrated in Ref.~\cite{Hatakeyama18},
the nuclear density distribution $\rho$ is well approximated
by the 2pF distribution $\rho_{\rm 2pF}$
with the nuclear radius $R$ and diffuseness $a$
parameters being fixed by minimizing
\begin{align}
  \int_0^\infty dr\, r^2|\rho(r)-\rho_{\rm 2pF}(r)|.
  \label{mindens.eq}
\end{align}
With this prescription,
the 2pF distribution nicely reproduces
the density profile near the nuclear surface
obtained from realistic mean-field calculations including
nuclear paring and deformation~\cite{Hatakeyama18}.
One can obtain the total diffuseness $a$ of
Eq.~(\ref{directa.eq}) by taking the differentiation
of $\bar{\rho}_i$ at $R$.
 I note that
the radius parameter $R$ in the 2pF distribution of Eq.~(\ref{2pF.eq})
is in general different
from the rms matter radius $R_m$ defined in Eq.~(\ref{radm.eq}).
An approximate relationship upto the second
order of $(a/R)^2$ is given as \cite{BM}
\begin{align}
  R_m^2\approx \frac{3}{5}R^2\left[1+\frac{7}{3}\pi^2\left(\frac{a}{R}\right)^2+
    \dots \right].
\end{align}
Apparently, $R_m=\sqrt{3/5}R$ holds if $a=0$ (sharp-cut radius) and
the nuclear diffuseness parameter $a$ plays a role to
enhance the nuclear radius $R_m$ from the reference radius of
the 2pF density $R$.

Figure~\ref{diff.fig} compares those obtained $a$ values
using Eqs.~(\ref{directa.eq}) and ~(\ref{mindens.eq}).
It should be noted that the relation of Eq.~(\ref{directa.eq})
is exact if the density distribution
is the 2pF function but in general it deviates from that.
Nevertheless, the $a$ values obtained from Eq.~(\ref{directa.eq})
are found to be close to the values obtained with
Eq.~(\ref{mindens.eq}): The deviation is at most 0.03 fm
and the square root of the rms deviation
between these diffuseness parameters is 0.014 fm for all isotopes
adopted in this paper.
This means that the surface region of those density distributions
are well approximated by the 2pF distribution,
and thus it makes sense to decompose the diffuseness parameters
into the s.p. contributions using the relation of Eq.~(\ref{directa.eq}).
In the following section, I discuss the
contribution of the s.p. orbit on the surface diffuseness in detail.

The averaged $a$ value is found to be 0.54 fm which is consistent with
the empirical one~\cite{BM,Chamon02}.
Even though the phenomenological
  Wood-Saxon parameter assumes the constant $a_{\rm WS}$ value,
  a variety of the $a$ values for the density distributions
  are obtained: Some kink structures at the magic and semi-magic numbers,
reflecting their s.p. density composition.
Detailed discussions will be given in the next section.
This paper aims to clarify the role of the s.p. wave function
in the total surface diffuseness.
I note, however, for more realistic cases,
the pairing interaction gives a smooth transition of the diffuseness
parameter across the magic numbers~\cite{Mizutori00, Horiuchi17,Hatakeyama18}.
Some kink structures disappeared owing to the mixing of the
s.p. levels near the Fermi level.
A detailed study with the pairing interaction
is interesting and worth studying in future
as an extension of this study.

\section{Results and discussion}
\label{results.sec}

\subsection{O isotopes}
\label{O.sec}

\begin{figure}[ht]
\begin{center}
  \includegraphics[width=0.49\linewidth]{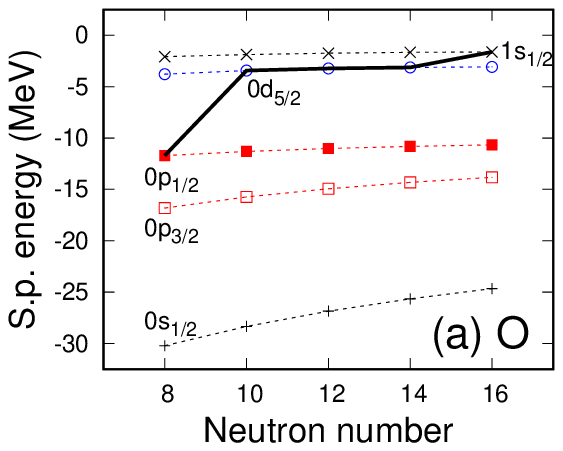}
\includegraphics[width=0.49\linewidth]{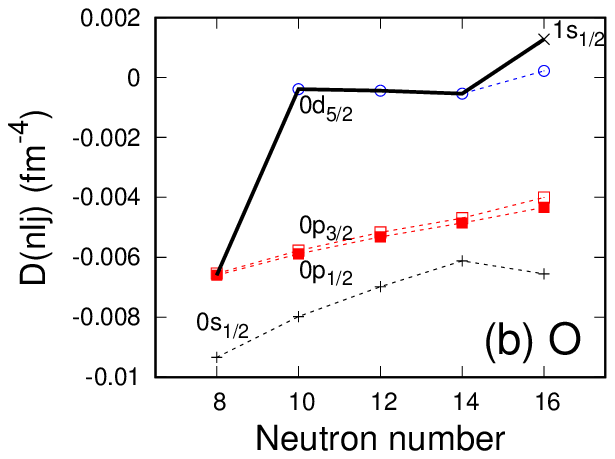}            
\caption{(a) Neutron single-particle (s.p.) energies
  and (b) derivative of the s.p. density at the radius parameter $R$
  for O isotopes. Thick black lines indicate the Fermi level, where
  the outermost neutrons are occupied.
  Plus and cross symbols indicate the $0s_{1/2}$ and $1s_{1/2}$ orbits,
  respectively.
  Squares and circles indicate the $p$ and $d$ orbits, respectively.
  Open and closed symbols distinguish the $j$-upper ($j_>=l+1/2$)
  and $j$-lower ($j_<=l-1/2$) orbits, respectively.
  Thin dotted lines are a guide to the eye.}
 \label{decomO.fig}
\end{center}
\end{figure}

Figure~\ref{decomO.fig} (a) displays
the neutron s.p. energies of $^{16-24}$O as a function of neutron number.
For a guide to the eye, a thick black line indicates
the Fermi level, where the outermost neutrons are occupied.
As seen in Fig.~\ref{diff.fig}, $^{16}$O has the smallest
diffuseness parameter among all the isotopes
due to well-bound $p$ orbits.
By filling neutrons in the $0d_{5/2}$ orbit near the Fermi level for $N>8$,
the surface diffuseness gradually increases
up to $N=14$. Finally, a sudden increase
of the surface diffuseness is obtained
due to the occupation of $1s_{1/2}$ orbit at $N=16$.

This behavior can be quantified and properly reflected in the derivative
of the s.p. orbit at the radius parameter $R$, i.e., $D_i$ of Eq.~(\ref{di.eq}).
For the sake of convenience, hereafter
the label of nucleon $i$ is denoted by the quantum numbers of
the s.p. wave function $nlj$, e.g., $D_i$ as $D(nlj)$.
Figure~\ref{decomO.fig} (b) plots the calculated $D(nlj)$ values
of each neutron s.p. orbit. The values are negative for
deeply bound $0s_{1/2}$, $0p_{3/2}$, and $0p_{1/2}$ orbits,
forming a sharper nuclear surface of $^{16}$O $(N=8)$, $a=0.460$ fm
compared to the standard value $0.54$ fm~\cite{BM}.
The $D(nlj)$ values for $0d_{5/2}$ orbit are almost zero.
This means that this s.p. orbit does not explicitly contribute
to changing the diffuseness parameter for $^{18-22}$O ($N=10$--14),
even though the $0d_{5/2}$ orbit
has the largest orbital angular momentum in O isotopes.
For $^{24}$O ($N=16$), since the $1s_{1/2}$ orbit has a positive $D(nlj)$ value,
the last two neutrons play a role in enhancing the surface diffuseness.

\begin{figure}[ht]
\begin{center}
  \includegraphics[width=0.49\linewidth]{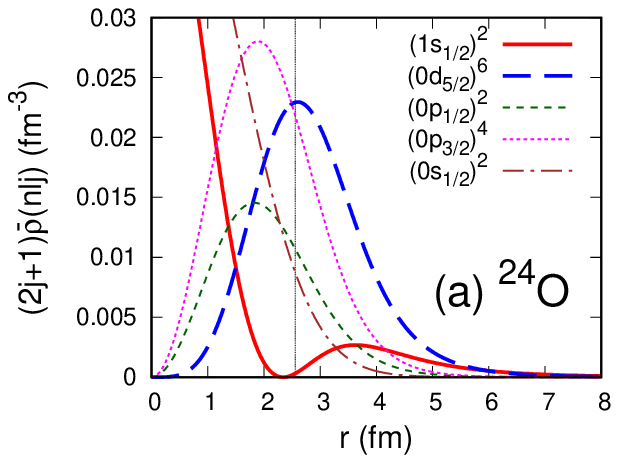}
  \includegraphics[width=0.49\linewidth]{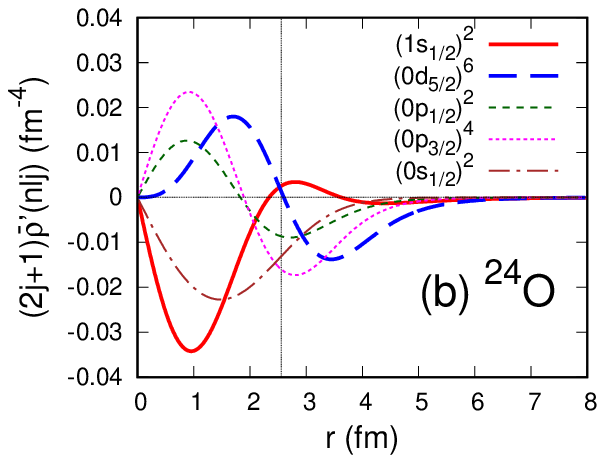}            
  \caption{(a) Neutron single-particle densities 
    and (b) their derivative of $^{24}$O multiplied by
    the occupation number $(2j+1)$.
    Vertical lines indicate the radius parameter $R=2.56$ fm of $^{24}$O
    and a horizontal line indicates zero.}
  \label{densO.fig}
\end{center}
\end{figure}

To understand the behavior of $D(nlj)$ intuitively,
Figs.~\ref{densO.fig} (a) and (b) respectively draw
the neutron s.p. densities $\bar{\rho}(nlj)$
and their derivatives
$d\bar{\rho}(nlj)/dr=\bar{\rho}^\prime(nlj)$ of $^{24}$O.
Each s.p. density is multiplied by
the number of occupied neutrons.
The radius parameter $R$ of $^{24}$O is indicated
as a vertical line in the plot.
For the deeply bound orbits with $0s_{1/2}$, $0p_{3/2}$, and $0p_{1/2}$,
the radius $R$ is located beyond these peak positions
as illustrated in Fig.~\ref{densO.fig} (a).
These $D(nlj)$ values are negative,
which can be seen quantitatively in Fig.~\ref{densO.fig} (b).
For $0d_{5/2}$ orbits, the $R$ value is located around
the peak position and thus the $D(nlj)$ value is almost zero.
A striking difference among the other orbits
is found in the $1s_{1/2}$ orbit. Because the $1s_{1/2}$ orbit has a node
and is further extended than the other orbits,
the $R$ position is located at an uphill towards the second peak
of the s.p. density, leading to a positive $D(nlj)$ value.

The global behavior of the s.p. contribution can also be understood
from Figs.~\ref{densO.fig} (a) and (b).
As seen in Fig.~\ref{decomO.fig} (b),
the absolute value of $D(nlj)$ for the deeply bound states
gradually increases because the radius parameter $R$
grows with increasing $N$.
Even though the ``valence'' $0d_{5/2}$ orbit does not explicitly contribute
to changing the surface diffuseness,
a moderate increase of the surface diffuseness is
found in $N=10$--$14$. This is mainly due to the change of
the nuclear surface defined by the radius parameter $R$.
Since the nuclear surface becomes
further from the surface of the ``core'' with increasing $N$,
in general, as expected from Fig.~\ref{decomO.fig} (b),
the s.p. orbits in the core, i.e., $0s_{1/2}$, $0p_{3/2}$, and $0p_{1/2}$,
orbits, contribute to enhancing the diffuseness parameter $a$.
The properties of the valence neutron orbits are
essential to explain the evolution of surface diffuseness.
Since the $1s_{1/2}$ orbit is extended and
enhances the nuclear matter radius drastically by
0.27 fm from $^{22}$O, which is about twice of
the enhancement in $N=10$--14,
the diffuseness $a$ and the radius parameter $R$ are
changed significantly at $N=16$, producing kinks of the $D(nlj)$ values
for the other s.p. orbits.

\subsection{Ca and Ni isotopes}
\label{CaNi.sec}

\begin{figure}[ht]
\begin{center}
\includegraphics[width=0.49\linewidth]{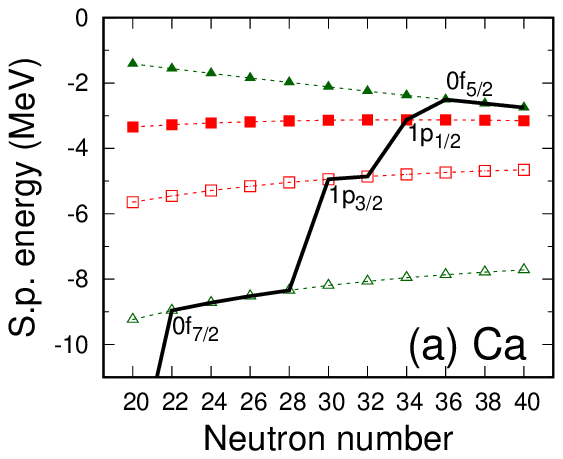}
\includegraphics[width=0.49\linewidth]{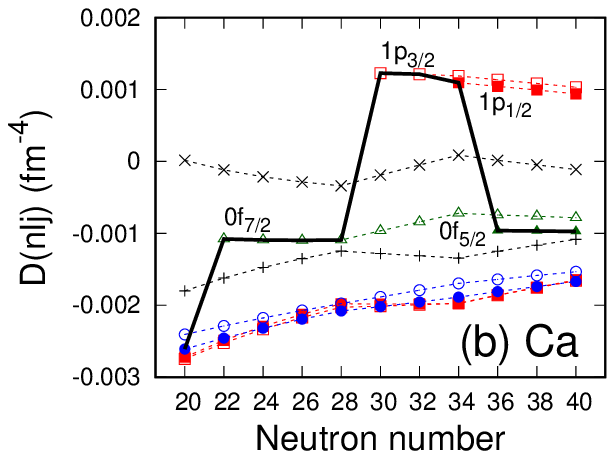}        
\caption{Same as Fig.~\ref{decomO.fig} but for Ca isotopes.
Only s.p. energies near the Fermi level are shown.
Triangles indicate the $f$ orbit.}
 \label{decomCa.fig}
\end{center}
\end{figure}

\begin{figure}[ht]
\begin{center}
\includegraphics[width=0.49\linewidth]{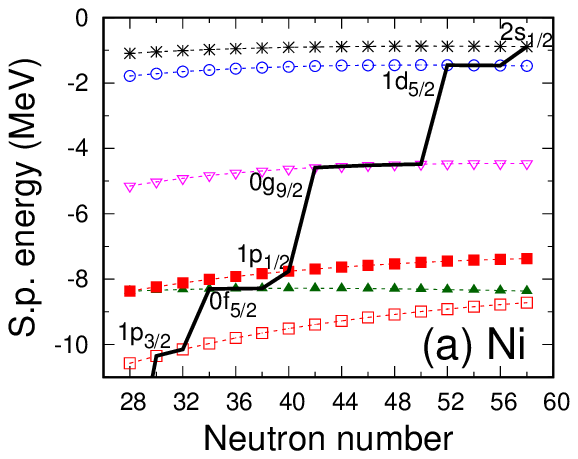}
\includegraphics[width=0.49\linewidth]{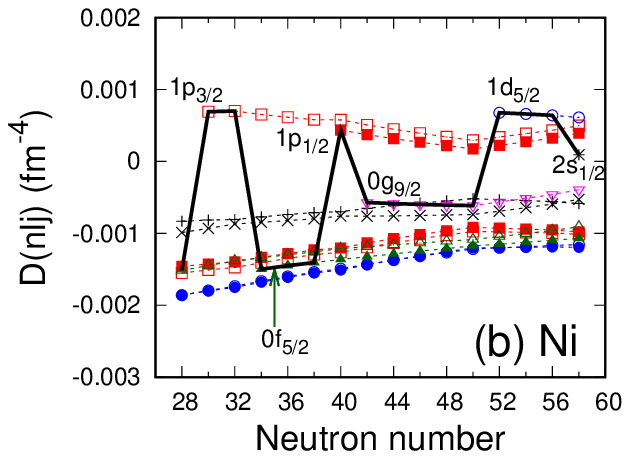}        
\caption{Same as Figs.~\ref{decomO.fig} and ~\ref{decomCa.fig}
  but for Ni isotopes.
  Only s.p. energies near the Fermi level are shown.  
  Inverted triangles and star symbols
  indicate the $g$ and $2s_{1/2}$ orbits, respectively.
}
 \label{decomNi.fig}
\end{center}
\end{figure}

Figures~\ref{decomCa.fig} and \ref{decomNi.fig} display the neutron
s.p. energies and $D(nlj)$ values of Ca and Ni isotopes
as a function of the neutron number.
Compared to O isotopes, which mainly consist of the nuclear surface,   
the absolute value of each $D(nlj)$ 
becomes smaller and the $N$ dependence becomes weaker.
Since the surface diffuseness consists of the sum of each s.p. contribution,
the weight of each s.p. contribution to the total
surface diffuseness becomes smaller for the medium-mass nuclei
compared to the light nuclei.
The isotope dependence of the surface diffuseness can easily be understood
by considering  the role of the ``core'' and ``valence'' neutrons.
For Ca and Ni isotopes, the core orbits, i.e.,
all the s.p. orbits below $N=20$ and 28, respectively,
gives similar $D(nlj)$ values due to deep binding and
contribute to gradually enhancing the surface diffuseness
as overall behavior\footnote{
At a closer look,
  similarly to the case of $0d_{5/2}$ of $^{24}$O in Fig.~\ref{densO.fig},
  the $D(nlj)$ value of the $1s_{1/2}$ orbit in Ca isotopes
  is almost zero, where the nuclear radius $R$ is
located at around the second peak of the s.p. density.
For Ni isotopes, since the $1s_{1/2}$ orbit is deeply bound
and the radius parameter $R$ is larger than Ca isotopes,
the $R$ value becomes being located beyond the second peak
of the s.p. density.},
while the valence neutron orbits
determine the evolution of the surface diffuseness,
depending on their quantum numbers

A filling of lower-$l$ or nodal orbit near the Fermi level
produces larger surface diffuseness
owing to a smaller centrifugal barrier.
In fact, as seen in Fig.~\ref{diff.fig} for Ca isotopes,
the diffuseness parameter at $N=20$--28 shows constant behavior
with relatively small diffuseness $\approx 0.5$ fm
because of a filling of the $0f_{7/2}$ orbit.
A sudden increase of the surface diffuseness occurs $N>28$
due to the occupation of the $1p$ orbits,
which is consistent with the implication from the recent interaction
cross-section measurement~\cite{Tanaka20} and theoretical
interpretation~\cite{Horiuchi20}.
The surface diffuseness stops growing due to the occupation
of the $0f_{5/2}$ orbit at $34<N\leq 40$.

\begin{figure}[ht]
\begin{center}
  \includegraphics[width=0.49\linewidth]{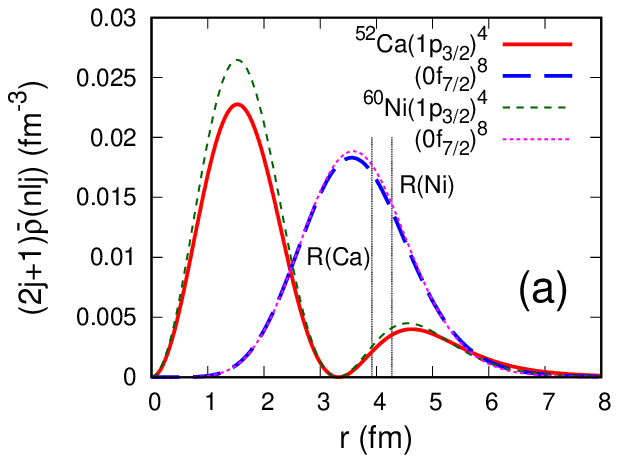}
  \includegraphics[width=0.49\linewidth]{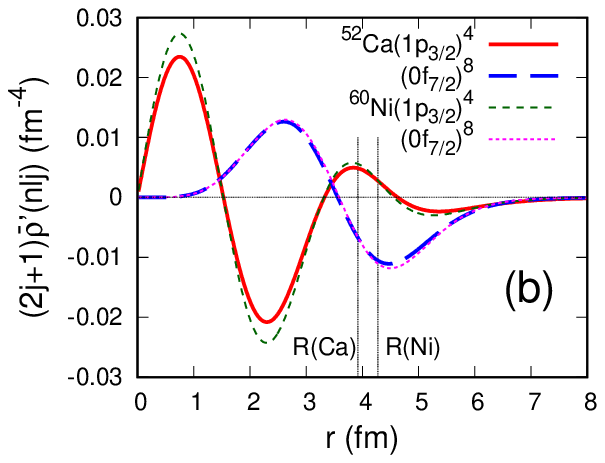}
  \caption{Same as Fig.~\ref{densO.fig} but for $^{52}$Ca and $^{60}$Ni.
    Single-particle states near the Fermi levels are selected.
    Vertical lines indicate the radius parameters,
    $R({\rm Ca})=3.92$ fm for $^{52}$Ca and
    $R({\rm Ni})=4.27$ fm for $^{60}$Ni.    
  }
 \label{densCaNi.fig}
\end{center}
\end{figure}

For Ni isotopes, in contrast to the case of the Ca isotopes
no drastic increase of the surface diffuseness is found at $N>28$.
In $28<N<40$, three sorts of s.p. orbits are filled in the order of
$1p_{3/2}$, $0f_{5/2}$, and $1p_{1/2}$.
This change of the filling orbit corresponds to the kinks at $N=32$ and 38.
For the same $N$, the surface diffuseness is smaller than the Ca isotopes.
The reason can be seen in Fig.~\ref{decomNi.fig},
which draws the neutron s.p. energies and $D(nlj)$ of Ni isotopes.
The $D(nlj)$ values of the $1p$ orbits are smaller than these of Ca isotopes.
This is partly because the $1p$ orbits are shrunk
due to deeper binding of the s.p. orbits in Ni isotopes and partly because
the radius parameter $R$ with the same $N$ is larger than that of Ca.
Figure~\ref{densCaNi.fig} plots the s.p. neutron densities
and their derivatives near the Fermi levels,
i.e., $0f_{7/2}$ and $1p_{3/2}$ of $^{52}$Ca and $^{60}$Ni.
The s.p. wave functions are shrunk, especially for the $1p_{3/2}$ orbit
of $^{60}$Ni compared to that of $^{52}$Ca.
The rms radius of the $0f_{7/2}$ and $1p_{3/2}$ orbits of $^{52}$Ca ($^{60}$Ni)
are 4.33 (4.26) and 4.76 (4.35) fm, respectively.
In addition to the shrinkage of the s.p. wave function,
the radius parameter $R$ is also larger for $^{60}$Ni.
Consequently,
the nuclear surface of $^{60}$Ni becomes sharper than $^{52}$Ca.
As seen in Figs.~\ref{decomCa.fig} (b) and \ref{decomNi.fig} (b),
the nodal $1p$ orbit always gives a
positive $D(nlj)$ value for Ni and Ca isotopes
because the radius parameter $R$ is
always located before the second peak of the s.p. density
as displayed in Fig.~\ref{densCaNi.fig}.

I remark that similar enhancement of the surface diffuseness
was found in neutron-rich Ne and Mg isotopes~\cite{Choudhary21}.
Interestingly, this occurs earlier than $N=28$ but $N>18$
in the so-called island of inversion,
where strong nuclear deformation is found.
The nuclear deformation allows the occupation of
the intruder orbit which induces configuration mixing
of the spherical $0f_{7/2}$ and $1p$ orbits.
Since the mixing includes the low-$l$ $1p$ orbits,
the surface diffuseness is drastically enhanced,
while the $0f_{7/2}$ mixing forms a sharper surface diffuseness.
This exemplifies that the discussion given in this paper remains
general for such deformed nuclei.
The nuclear nuclear deformation in general induces
fractional occupation numbers near the Fermi level,
which may include lower-$l$ orbits,
leading to larger surface diffuseness~\cite{Horiuchi21}.

\begin{figure}[ht]
\begin{center}
  \includegraphics[width=0.49\linewidth]{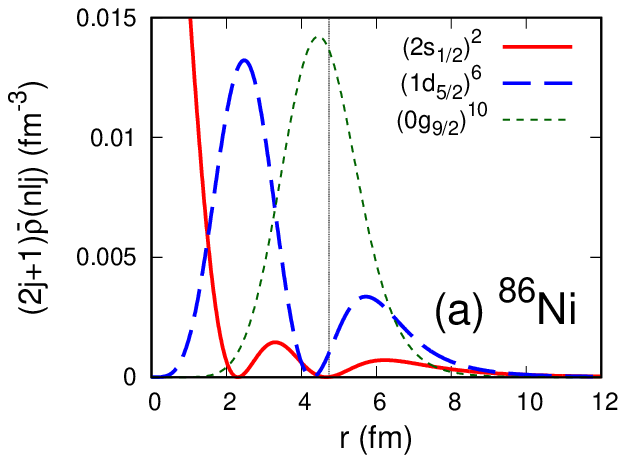}
  \includegraphics[width=0.49\linewidth]{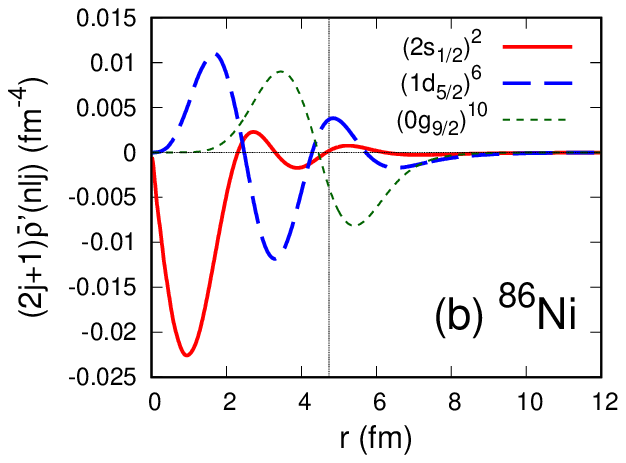}  
  \caption{Same as Fig.~\ref{densCaNi.fig}
    but for $^{86}$Ni.
    Vertical lines indicate the radius parameter $R=4.74$ fm.
    }
 \label{dens86Ni.fig}
\end{center}
\end{figure}

A sudden increase of the surface diffuseness occurs at $N>50$
after ``sharp'' $0g_{9/2}$ orbits are fully occupied
for Ni isotopes due to the occupation of
the lower-$l$ orbits, $2s_{1/2}$ and $1d_{5/2}$.
For $^{86}$Ni, as seen in Fig.~\ref{decomNi.fig} (b), 
the nuclear surface is diffused mostly by the filling of the $1d_{5/2}$ orbits.
Figure~\ref{dens86Ni.fig} plots these s.p. densities
and their derivatives for $^{86}$Ni.
The radius parameter $R$ is located inside
the second and third peaks for $1d_{5/2}$ and $2s_{1/2}$ orbits,
respectively, giving positive $D(nlj)$ values, while the nodeless
$0g_{9/2}$ orbit gives a negative
derivative of the s.p. density at the radius parameter $R$
because the $R$ value is located just after the peak
of the s.p. density.
Note that in Fig.~\ref{decomNi.fig} (b)
the $D(nlj)$ value of the $2s_{1/2}$ orbit is smaller
than that of the $1d_{5/2}$ orbit which has higher angular momentum.
This is partly because the $s$ wave has no centrifugal barrier
and thus smaller derivative of the s.p. density in
the nuclear surface region is in general obtained compared to $1d_{5/2}$ orbit,
and partly because the $R$ value is
accidentally located at the second dip of the s.p. density.
For this $^{86}$Ni case, though the $2s_{1/2}$ orbit
does not contribute to changing the diffuseness parameter,
it plays a role in forming a halo tail
as the s.p. energy is small $-$0.888 MeV, resulting in
the rms s.p. radius 7.60 fm, which is much larger
than the matter radius of $^{86}$Ni, 4.56 fm.

\subsection{Sn isotopes}
\label{Sn.sec}

\begin{figure}[ht]
\begin{center}
\includegraphics[width=0.49\linewidth]{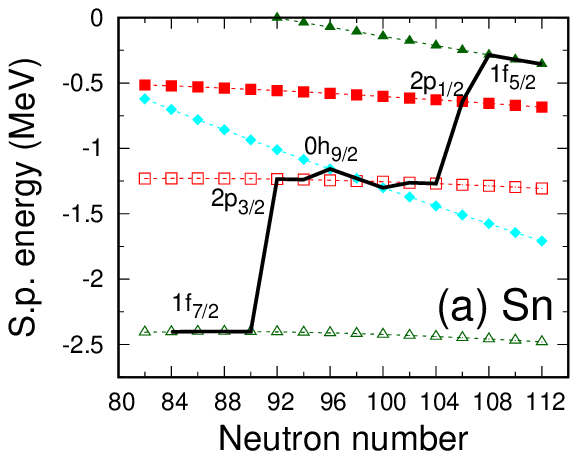}
\includegraphics[width=0.49\linewidth]{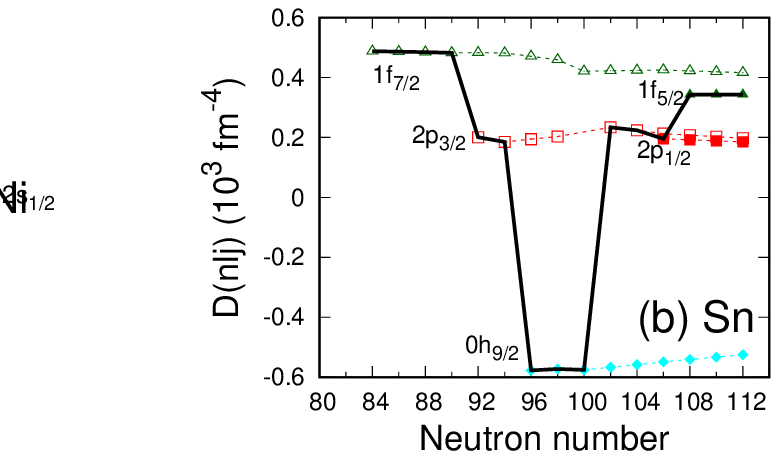}        
\caption{Same as Figs.~\ref{decomO.fig},
    \ref{decomCa.fig}, and \ref{decomNi.fig}
    but for Sn isotopes.
    Only s.p. energies near the Fermi level and
    corresponding $D(nlj)$ values are shown.    
    Closed diamonds indicate the $0h_{9/2}$ $(l=5)$ orbit.
Note that the $D(nlj)$ values are multiplied by 10$^3$ for the sake of convenience.}
\label{decomSn.fig}
\end{center}
\end{figure}

As was shown in Fig.~\ref{diff.fig} for Sn isotopes,
no strong enhancement of the surface diffuseness occurs in $50<N<56$.
This can be understood by the same reason found
in the comparison of Ca and Ni isotopes in $28<N<40$.
At $N=50$--70, the $1d_{5/2}$, $0g_{7/2}$, $1d_{3/2}$, and $2s_{1/2}$
orbits, which belongs to the principal quantum number $q=2n+l=4$,
are contributed and change the surface diffuseness:
The nodal s.p. orbit diffuses the nuclear surface,
while the nodeless s.p. orbit sharpens it.
At $N=70$--82, the reduction of the surface diffuseness
is due to the occupation of the $0h_{11/2}$ orbit ($q=5$).

\begin{figure}[ht]
\begin{center}
  \includegraphics[width=0.49\linewidth]{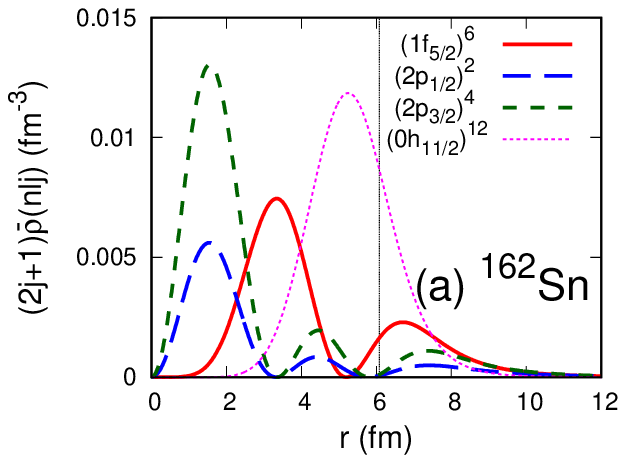}
  \includegraphics[width=0.49\linewidth]{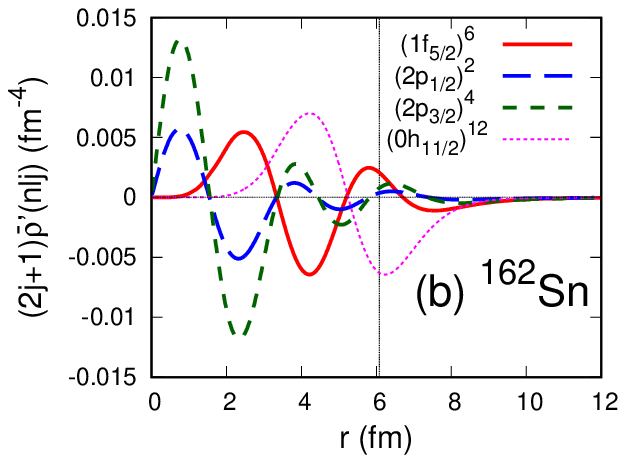}  
  \caption{Same as Fig.~\ref{densCaNi.fig}
    but for $^{162}$Sn.
    Vertical lines indicate the radius parameter $R=6.08$ fm.
  }
  \label{densSn.fig}
\end{center}
\end{figure}

Figure~\ref{decomSn.fig} plots the s.p. energies and $D(nlj)$ values for $N>80$.
A sudden increase of the surface diffuseness $N>82$
is due to the occupation of the nodal $1f_{7/2}$ orbit ($q=5$).
Level inversions of the low-$l$ $2p_{3/2}$
and sharp $0h_{9/2}$ orbits occurs which produces the wiggles
at $N\approx 94$--100.
It is again noted that, in reality, these kink structures disappear
because those s.p. orbits are mixed by the pairing
correlation~\cite{Hatakeyama18},
which induces fractional occupation of the orbits near
the Fermi level. 
At $N>100$, strong enhancement of the surface diffuseness is found.
Regarding $^{150}$Sn as a core nucleus,
all the nodal valence orbits with
$1f_{7/2}$, $2p_{3/2}$, $2p_{1/2}$, and $1f_{5/2}$,
have positive $D(nlj)$ values and contribute 
to increasing the surface diffuseness.
Figure~\ref{densSn.fig} displays the s.p. densities
and their derivatives of $^{162}$Sn.
While the outermost core orbit $0h_{11/2}$ gives
a negative $D(nlj)$ value, all these nodal s.p. orbits show
positive $D(nlj)$ values which form the diffused nuclear surface.
The higher $l$ value, the larger magnitude of the $D(nlj)$ value
becomes because the variation of the s.p. density
near the nuclear surface becomes large for the high-$l$ state.

\subsection{Pb isotopes}
\label{Pb.sec}

\begin{figure}[ht]
\begin{center}
\includegraphics[width=0.48\linewidth]{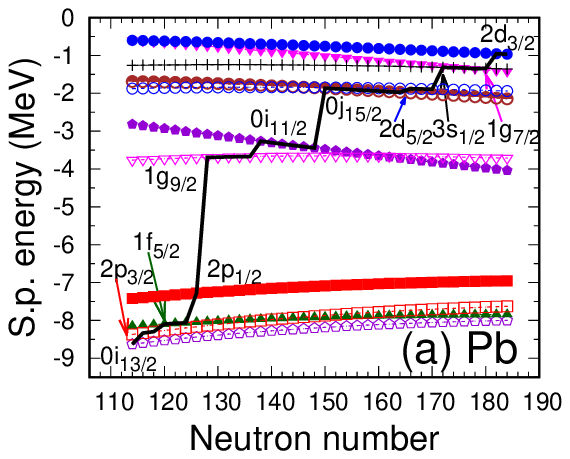}
\includegraphics[width=0.48\linewidth]{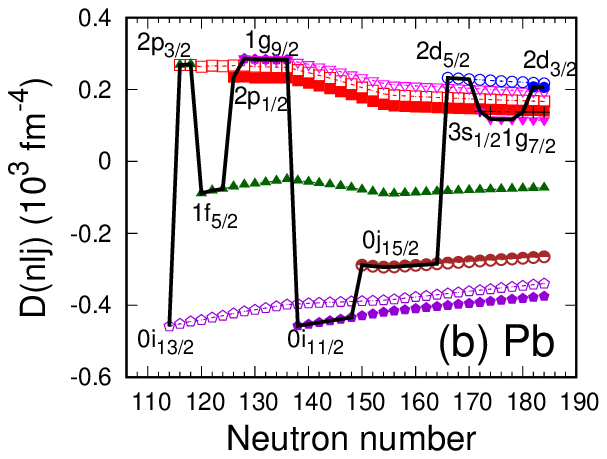}        
\caption{Same as Figs.~\ref{decomO.fig},
  \ref{decomCa.fig}, \ref{decomNi.fig}, and \ref{decomSn.fig}
  but for Pb isotopes.
  Pentagons, half-open circles, and plus symbols
  indicate the $0i$ ($l=6$), $0j_{15/2}$ ($l=7$), and $3s_{1/2}$ orbits,
  respectively.
    Only s.p. energies near the Fermi level and
    corresponding $D(nlj)$ values are shown.      
Note that the $D(nlj)$ values are
  multiplied by 10$^3$ for the sake of convenience. 
}
 \label{decomPb.fig}
\end{center}
\end{figure}

Finally, I discuss Pb isotopes. As seen in Fig.~\ref{diff.fig},
I find apparent kink behavior at the neutron magic numbers,
where the nuclear major or subshell is fully occupied.
However, $N=126$ is one exception that no prominent
kink behavior is found even though the major shell
is fully occupied. I remark that a similar result
was already shown in Ref.~\cite{Hatakeyama18}.
Figure~\ref{decomPb.fig} plots the neutron s.p. energies
and $D(nlj)$ values of Pb isotopes.
At $114<N \leq 126$ the surface diffuseness grows
by filling neutrons in $2p$ and $1f_{5/2}$ orbits $(q=5)$.
For $N>126$, the major shell is changed to $q=6$; however,
the nodal $1g_{9/2}$ orbit, which has a positive $D(nlj)$ value,
leads to a further increase of the surface diffuseness
and thus the kink structure disappears.

\begin{figure}[ht]
\begin{center}
  \includegraphics[width=0.49\linewidth]{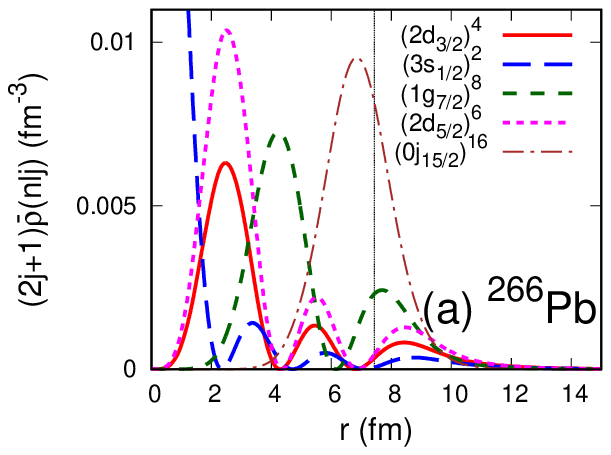}
  \includegraphics[width=0.49\linewidth]{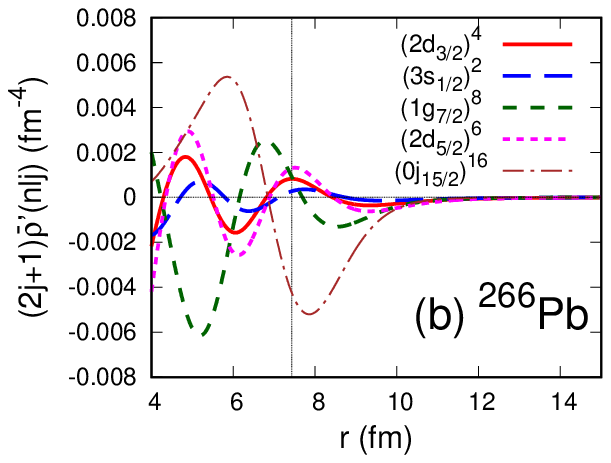}          
  \caption{
    Same as Fig.~\ref{densCaNi.fig}
    but for $^{266}$Pb.
    Vertical lines indicate the radius parameter $R=7.43$ fm.
    Note that the values $r<4$ are not shown in the panel (b)
    for the sake of visibility.
  }
  \label{densPb.fig}
\end{center}
\end{figure}

After the $1g_{9/2}$ orbit is fully occupied,
the surface diffuseness decreases at $136<N\leq 164$
because neutrons occupy the nodeless $0i_{11/2}$ and $0j_{15/2}$ orbits,
which have a fast drop-off of the wave function
at the radius parameter $R$.
At $N>164$, the diffuseness parameter is again enhanced by the filling
of the nodal s.p. orbits $2d_{5/2}$, $3s_{1/2}$, $1g_{7/2}$ and $2d_{3/2}$
up to $N=184$.
Figure~\ref{densPb.fig} plots the neutron s.p. densities
and their derivatives of $^{266}$Pb.
The occupation of and the outermost core ($^{246}$Pb) orbit,
$0j_{15/2}$, forms shape nuclear surface
but the other 20 nucleons in these $2d_{5/2}$, $3s_{1/2}$, $1g_{7/2}$,
$2d_{3/2}$ orbits produce the diffused nuclear surface
beyond $N=164$. The $l$ dependence of these nodal s.p. orbits are small.
Their individuality is almost lost for such a heavy nucleus.

\subsection{Characteristics of s.p. densities}
\label{global.sec}

\begin{figure}[ht]
\begin{center}
\includegraphics[width=1\linewidth]{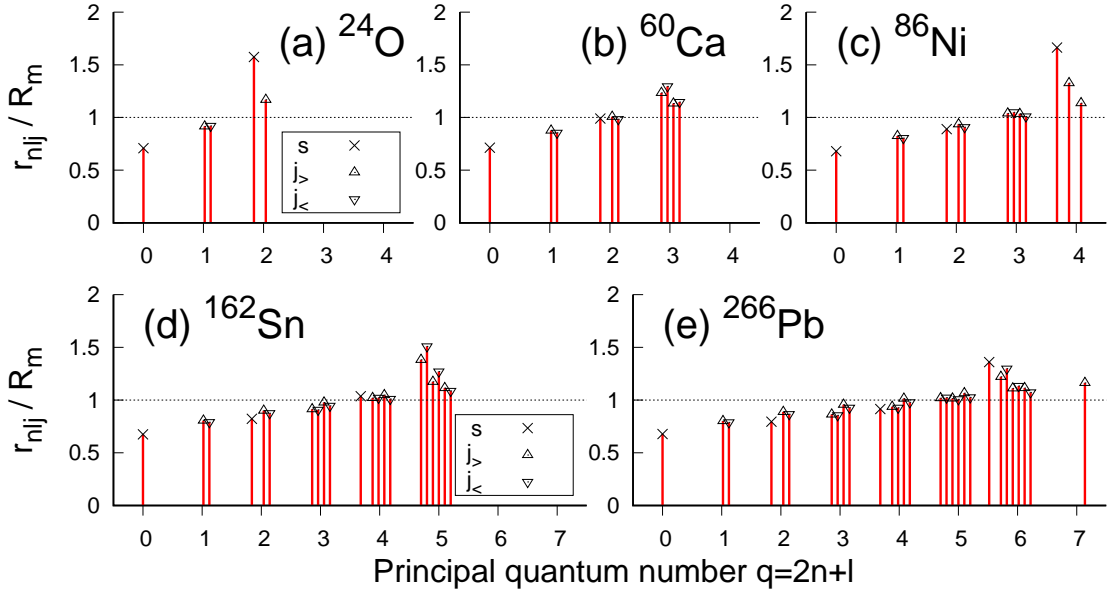}        
\caption{Rms radius of the neutron s.p. density
  divided by the rms matter radius
  categorized into the principal quantum number $2n+l$.
  Crosses, triangles, and inverted triangles indicate
  $s$, $j_>=l+1/2$, and $j_<=l-1/2$ orbits, respectively.
  The values in the same $q=2n+l$
  are arranged from left to right in the order of
  $j_>$ and $j_<$ from low to high $l$.
  Bars are a guide to the eye.  
  Horizontal dotted lines indicate unity.
}
\label{spradii.fig}
\end{center}
\end{figure}

Finally, I discuss why the nodal or lower-$l$ s.p. orbits
near the Fermi level are important to form a diffused nuclear surface.
It is known that the rms radius
of the harmonic oscillator (HO) s.p. wave function depends
only on the principal quantum number $(q=2n+l)$.
The difference appears when more realistic mean-field potential
is considered. The potential in general includes
diffused central and surface-type spin-orbit potentials,
resulting in the enhancement of the amplitude of
the s.p. wave functions near the nuclear surface,
especially for the s.p. states near the Fermi level.

Figure~\ref{spradii.fig} plots
the neutron rms s.p. radii divided by the rms matter radius
for the neutron dripline nuclei predicted by the present parameter set, i.e.,
$^{24}$O, $^{60}$Ca, $^{86}$Ni, $^{162}$Sn, and $^{266}$Pb,
categorized into the principal quantum number $q=2n+l$.
The maximum principal quantum number $q_{\rm max}$ is 
2, 3, 4, 5, and 6--7 for $^{24}$O, $^{60}$Ca, $^{86}$Ni, $^{162}$Sn, and $^{266}$Pb,
respectively.
For the sake of convenience, the results in the same $q$
are arranged from left to right
in the order of $j_>=l+1/2$ and $j_<=l-1/2$ from low to high $l$,
for example, the order of $1s_{1/2}, 0d_{5/2}$, and $0d_{3/2}$ for $q=2$;
and $1p_{3/2}, 1p_{1/2}, 0f_{7/2}$, and $0f_{5/2}$ for $q=3$.

For deeply bound or ``core'' orbits
$q\leq q_{\rm max}-1$ ($q\leq 5$ for $^{266}$Pb),
the rms s.p. radii are roughly proportional to $q=2n+l$,
i.e., almost constant behavior of the rms neutron s.p. radii
is found with respect to $l$ and $j$
at the same major shell as expected from the properties
of the HO wave function.
Since these deeply bound states have smaller rms radii than
the nuclear radius, the $D(nlj)$ value is negative 
or at most small positive value, which
does not cause the sudden increase of the surface diffuseness but
it induces a gradual increase of the surface diffuseness on $N$.
In contrast to this, the rms s.p. radii that belong to
the maximum principal quantum number $q_{\rm max}$
exhibit some angular momentum dependence: The lower $l$, the larger
rms s.p. radius becomes.
The behavior of the rms neutron radii near the Fermi level
can be explained simply by considering the
change of the classical turning point on the effective potential well.
The lower $l$, the larger rms radius becomes
because the lower-$l$ orbits have in general
more penetrability at the surface region if the binding s.p. energy is the same.
For the states near the Fermi level,
the radius of the classical turning point on the effective potential well
becomes larger due to the surface-diffused nuclear potential,
leading to a large rms s.p. radius.
This also lowers the centrifugal barrier and
further increases the rms s.p. radius.
The $j_<$ orbit gives a smaller rms radius
than that of $j_>$ one because the
repulsive spin-orbit interaction near the nuclear surface
which acts opposite to the $j_>$ state.

This non-trivial increase of the rms s.p. radius
near the Fermi level is essential to explain
the evolution of the nuclear surface diffuseness.
In Fig.~\ref{spradii.fig},
the rms s.p. radii that belong to $N_{s,{\rm max}}$ exceed
unity, i.e., that exceed the rms matter radius.
The $D(nlj)$ value of the nodal s.p. orbits near the Fermi level
can be positive, in which the radius parameter $R$ is located
before the last peak position, while the $D(nlj)$ value is negative
for the nodeless high-$l$ s.p. orbit.
Deducing the surface density will be important for detailed
study of the nuclear s.p. orbits near the Fermi level.
For example, interesting modification
of the s.p. wave functions due to the spin-orbit interaction
near the Fermi level was suggested in Ref.~\cite{Nakada19},
which can be essential for explaining a puzzle of the kink structure
of the charge radii of Pb isotopes.

\section{Conclusion}
\label{conclusion.sec}

To extract spectroscopic information
from the nuclear surface diffuseness,
I have proposed a practical and convenient way to decompose
the surface diffuseness into contributions of each
single-particle (s.p.) orbit.
The nuclear surface diffuseness defined in a familiar two-parameter
Fermi density distribution is inversely proportional to
the sum of the derivatives
of the s.p. wave functions at the nuclear surface radius.
I have quantified its contributions
to the surface diffuseness for spherical
neutron-rich O, Ca, Ni, Sn, and Pb isotopes
using a phenomenological mean-field model.

I find that the neutron number dependence of the surface diffuseness
can simply be understood by the quantum number of the occupied
s.p. orbits near the Fermi level:
The occupation of nodeless s.p. orbits induces
a mild change of the surface diffuseness,
while the occupation of nodal s.p. orbits 
enhances the surface diffuseness
because the nodal s.p. orbit near the Fermi level can have relatively
large s.p. radius that exceeds the matter radius
and lower the orbital angular momentum.
The enhancement becomes significant when the neutron dripline is approached
where a sudden increase of the s.p. radius near the Fermi level is expected.

The present method can be applied to
more realistic nuclear systems 
that include configuration mixing of spherical s.p. orbits
induced by, e.g., nuclear deformation and pairing correlations,
if the decomposition of their s.p. orbits into
the spherical s.p. orbits is made.
Applied to these systems, it may be possible to quantitatively
show how the density distribution near the nuclear surface
is composed microscopically,
depending on the degree of these correlations.
It should be noted that
the surface diffuseness is a robust physical quantity
that can be measured by, e.g., proton elastic
scattering~\cite{Hatakeyama18,Choudhary20,Choudhary21}.
A systematic measurement of the surface diffuseness
strongly facilitates the understanding of various phenomena
near the nuclear surface. 

\section*{Acknowledgment}

This work was in part supported by JSPS KAKENHI Grants No. 18K03635.
I acknowledge the collaborative research program 2021, 
Information Initiative Center, Hokkaido University.

\end{document}